\documentclass[pra,twocolumn,aps,preprintnumbers]{revtex4}
\usepackage{graphicx}

\begin{document}

\title{Programmable Logic Devices in Experimental Quantum Optics}
\author{J. Stockton}
\email{jks@caltech.edu}
\author{M. Armen}
\author{H. Mabuchi}
\affiliation{Norman Bridge Laboratory of Physics 12-33, California
Institute of Technology, Pasadena, California 91125 USA}

\date{\today}

\begin{abstract}
We discuss the unique capabilities of programmable logic devices
(PLD's) for experimental quantum optics and describe basic
procedures of design and implementation.  Examples of advanced
applications include optical metrology and feedback control of
quantum dynamical systems. As a tutorial illustration of the PLD
implementation process, a field programmable gate array (FPGA)
controller is used to stabilize the output of a Fabry-Perot
cavity.
\end{abstract}

\maketitle

\section{Introduction}

Automatic controllers are pervasive in experimental physics. Servos typically
play a role behind the scenes, stabilizing environmental conditions ({\it
e.g.} temperature, frequency and amplitude of driving lasers) for the
physical system of primary interest ({\it e.g.} quantum dots, trapped atoms
or molecules). But the system of interest can itself be the explicit object
of sophisticated control strategies. An increasing number of experimental
quantum systems are developing to the point where coherent dynamics occur at
a time scale longer than that of available detectors and actuators
\cite{ion,acm,bec}. This separation of time scales opens the door for
real-time feedback control to be applied in quantum-mechanical scenarios.

New theoretical and experimental tools will be required to achieve quantum
control objectives. Concerted efforts are currently being made to extend
classical control theory to quantum problems where back-action cannot be
ignored \cite{doherty,adphi}. Given the inherent nonlinearity of conditional
quantum dynamics, optimal control laws cannot be practically implemented with
analog circuits, necessitating fast digital control. Even for linear systems,
programmable logic may be superior to analog methods when a precisely shaped
transfer function is desired. For these reasons, one expects that
programmable logic devices (PLD) with high processing speed and low latency
will prove to be invaluable as quantum and classical controllers.

PLD's are already a standard tool in industry and some areas of science, but
they have yet to attain widespread use in fields such as quantum optics and
quantum information science. Our aim in this paper will be to convey a base
level of knowledge required to use these devices in representative
experimental setups. First, we motivate the use of programmable logic with
some potential applications. We then describe the details of practical
implementation, from determining the required hardware specifications to
completing the design flow. Finally, we demonstrate this process with a
familiar example of classical optical control by using a Field Programmable
Gate Array (FPGA) to lock a Fabry-Perot cavity.

\section{Applications}

An outstanding feature of PLD's is that they can implement complex non-linear
logic with relatively low latency. Here `latency' refers to the delay between
the time that a signal is received as input and the time that a calculation
based on it becomes available as output. This reaction time is of little
consequence in many data-processing applications, but is critical in control
loops. The control bandwidth of any servo is limited by the
inverse of this delay.

In addition, most PLD's can be completely re-programmed in
a matter of minutes, allowing for a high degree of design
flexibility in experimental situations. Given a PLD with these
capabilities, it is not difficult to imagine a variety of control
applications related to quantum optics. Here we summarize a few
potential examples, some of which are currently being developed.

\subsection{Precise linear servos}

In linear control tasks, PLD controllers have a distinct practical
advantage over analog circuitry with regard to precision and
flexibility. For example, it is a well known control problem to
stabilize a plant over one of its resonances. An appropriate
controller should precisely compensate the measured center
frequency and quality factor of the resonance. When creating an
analog servo the designer must work with discrete components
(resistors, capacitors, etc.) whose impedances have a
non-negligible error range. However a PLD transfer function can be
specified digitally, making it much easier to closely match the
system dynamics.

\begin{figure}
\includegraphics[width=3in]{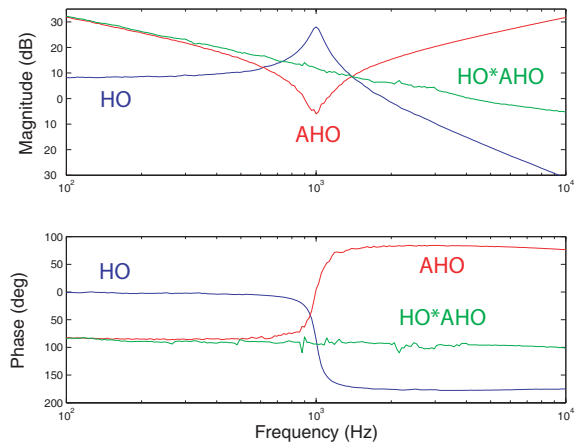}
\caption{The blue plot is a harmonic oscillator (HO) transfer
function and the red plot is the anti-harmonic-oscillator (AHO)
transfer function. The product of the two should resemble an
integrator transfer function (green) with a constant -90 degree
phase. } \label{aho}
\end{figure}

Figure \ref{aho} shows the near-compensation of a harmonic
oscillator (HO) resonance with a PLD `anti-harmonic-oscillator'
(AHO) transfer function. (Actually, both transfer functions in the
graph are implemented with a PLD by techniques described later.)
Ideally, the HO transfer function will be transformed into an
integrator transfer function (with a constant -90 degrees of
phase) when multiplied by the AHO compensator.  The deviation from
a perfect integrator is due to a slight error in the assumed
damping. Refinements to the AHO design could remove this
non-ideality.

PLD's will obviously not replace every linear servo in the typical
laboratory, but the ability to optimize the stability of critical
laser systems (for example) is a considerable resource. We detail
the use of a PLD controller to optimally perform a linear control
task in a later section.

\subsection{Optimal measurement}

In quantum feedback scenarios, either the measurement operators or
the system Hamiltonian can be modulated in real time according to
the information gained from a continuous measurement record.

Consider the case where only the measurement operators are
adjusted. The goal of the entire measurement may be to most
accurately determine the initial state of the system.  Other
situations may call for the measurement of only a single state
parameter, where all other state variables are either assumed or
neglected. The authors are currently developing a system of this
type where the goal is to optimally measure the phase of a
\textit{single} pulse of light. We constrain ourselves to
measuring pulses that are long enough to have their phase be well
defined and also long enough to allow us to feedback the
measurement signal multiple times before the pulse has been
completely destroyed by the detectors.

Wiseman \emph{et al.} have determined close-to-optimal measurement
schemes for this system based on quantum trajectory theory
\cite{adphi}. In short, they consider the signal to be measured in
an adaptive homodyne set-up where the pulse is mixed with a strong
local oscillator whose phase, $\Phi$, is continuously adjusted
(within the duration of each pulse) according to the measured
homodyne current, $I$. To first order, the job of the algorithm is
to lock to the side of the interference fringe, thus $\Phi$ is
adjusted until $I$ is zero.

Despite this simplistic description, the general optimal algorithm
($f:I\Rightarrow \Phi $) is a highly non-linear function based on state
estimation. It has been shown that the estimated state at any time is a
function of only two parameters and the initial conditions. In terms of a
scaled time $v$, the parameters are
\begin{eqnarray}
A_{v}=\int_{0}^{v}I(u)e^{i \Phi(u)}du \\
B_{v}=-\int_{0}^{v}e^{2i\Phi(u)}du
\end{eqnarray} The phase of the local oscillator is usually taken to be $\Phi(v)=
\hat{\phi}(v)+\frac{\pi}{2}$ where $\hat{\phi}(v)$ is the phase
estimate to be used during the course of feedback. If one were to
stop the feedback at any time, the best phase estimate would be
$\hat{\phi}_{C}(v) = arg(C_{v})$ where
$C_{v}=A_{v}v+B_{v}A_{v}^{*}$. However, for subtle reasons,
$\hat{\phi}_{C}(v)$ should not be used as the estimate during the
course of the feedback.

One simple algorithm uses $\hat{\phi}(v)=arg(A_{v})$. With this choice, the
algorithm simply reduces to a gain-scheduled integrator of the form
\begin{equation}
d\Phi(v)=\frac{I(v)}{\sqrt{v}}
\end{equation}
where $v$ is the time since the beginning of the pulse and the $\sqrt{v}$
factor represents the effective gain.  Currently, this algorithm is being
implemented with an FPGA that creates the $\sqrt{v}$ gain factor with a
look-up table representation of the function as described in a later section.

More sophisticated algorithms (with optimal performance for certain squeezed
states) have been proposed that use feedback of the form
\begin{equation}
\hat{\phi}(v)=arg(C_{v}^{1-\epsilon(v)}A_{v}^{\epsilon(v)})
\end{equation}
where $\epsilon(v)$ is also a function of $A_{v}$ and $B_{v}$. In
this case, the algorithm is sufficiently complex that any analog
implementation would be extremely difficult to design.

In any case, the non-linear, low latency behavior of PLD's suggest that they
are a suitable tool for this task. Given that the form of a desired algorithm
may change frequently with the introduction of realistic experimental
complications, the rapid prototyping allowed by a PLD is also extremely
convenient.

\subsection{Feedback control}

When the goal is control rather than optimal measurement, a non-trivial
Hamiltonian of the system will be controlled by the measurement record.
Consider the case of an atom drifting through the light field of a small
Fabry-Perot cavity. As has been demonstrated, the position of the atom may be
imprinted onto the output light of the cavity \cite{acm}. This information
can potentially be mapped back onto the intensity and phase of the input
laser with the goal of trapping the atom in the cavity for extended periods
of time \cite{trap}.

Optimal control of the atom's position will require a complex
predictor-corrector structure in the feedback loop at $\mu$sec time scales.
If the associated calculations can be sufficiently reduced, a PLD with
effective clocking speeds above a MHz will be able perform this task. Of
course, the effectiveness of the control algorithm will depend on the assumed
dynamics of the system from which it is derived. If the system needs to be
described quantum mechanically, we should institute a conditional quantum
state estimator. If a classical description is sufficient, we can use a less
complicated algorithm. The performance of different controllers will be a
strong indicator of the validity of our descriptions.  The ability to quickly
re-design the PLD will be particularly advantageous when exploring this
boundary.

Hamiltonian feedback can also be used to manipulate the internal states of
atomic and molecular systems.  Numerous groups have become interested in
shaping femtosecond laser pulses to drive transitions which may be
inaccessible using traditional means \cite{chem}.  This includes the ability
to synthesize rare molecular compounds.  For example, by iteratively reading
the fluorescence spectrum of the system and intelligently moving in the
parameter space of the pulse shape, one attempts to land at a shape conducive
to creating the desired state or compound.

This procedure can happen in two regimes, `learning control' or `feedback
control'.  For learning control we consider using a new sample for every
pulse, whereas for feedback control we consider using the same sample on
every pulse. In the latter case, the algorithm assumes that the sample has a
long enough dephasing time (memory) that a significant degree of coherence is
retained between pulses. For either case, especially the second, a PLD based
controller may have significant advantages over alternative controller
architectures.

\subsection{Decision and control for quantum information processing}

In a generic quantum computing architecture, there exist classical logic
steps which involve performing a coherent quantum operation conditioned on
the result of a measurement. For example, quantum error correcting codes can
combat decoherence by mapping measured errors to appropriate correction
operators \cite{book}. In an experiment, this measurement-operation procedure
should be performed much faster than the dephasing rate of the system. If the
operations can be performed quickly upon command, PLD's will be able to
orchestrate these codes in a reliable and reconfigurable fashion with minimal
delay.

Even for non-conditional algorithms, PLD's can streamline the implementation
of complex instruction sets.  In particular, groups working on ion trap
computing have developed means of performing entanglement algorithms
\cite{ion}, but with an extensive overhead of macroscopic equipment that
requires detailed manual adjustment whenever the algorithm is changed.
Without pushing its computational limits, a PLD can be made to streamline
such logic networks. By using software defined algorithms, the users
eliminate the time and risk of error associated with manual realignment of
network components.  Commercial magnetic resonance systems use PLD's for
similar reasons.

As quantum computing architectures grow to the point where conditional and
non-conditional algorithms must be integrated in a way that is fast and
flexible, programmable logic will be able to handle the task in a convenient
manner.

The success of any PLD controller will depend on its dynamic range and
effective bandwidth. Next we discuss in more practical terms what levels of
system performance can be reasonably expected from currently available PLD's.

\section{Design}

\subsection{Hardware}

Once it is determined that a control algorithm needs to be
implemented digitally, a designer is confronted with a wide array
of possible controllers and corresponding acronyms.  In addition
to PLD's, the options include conventional microprocessor systems,
DSP's (digital signal processors), and ASIC's (application
specific integrated circuits). Of course, the choice of controller
is highly dependent on the algorithm being implemented because
each device has its own trade-offs. Microprocessor systems are
general enough to allow for a simple means of programming complex
algorithms. However, these systems rely on a single bus
architecture which forms a significant bottleneck in signal
processing applications. Overall throughput may be high, but a
large delay limits typical controllers to slow applications with
kHz scale bandwidths.  In addition, unreliable operating systems
may present undesirable interrupt signals during critical stages
of processing. DSPs are specialized microprocessor systems with a
multiple bus design that are optimized for signal processing
applications. Due to their parallel architecture, DSP's can attain
low-latency performance, but require a significant degree of
high-level design expertise. ASIC's are like PLD's in that the
user designs them from the gate level, but ASIC's are irreversibly
hardwired with a single application. While PLD's generally have
fewer resources available than ASIC's, they offer an efficient
parallel computation structure along with reprogrammability and a
relatively simple design process \cite{dsp}.

The market for PLD's is currently dominated by two companies:
Xilinx and Altera.  Devices from both companies have had extensive
product development in industry, thus a substantial support
network is available to designers. In choosing between PLD
companies, several factors beyond the chip performance need to be
considered, including the quality of the associated software
environments.  To obtain the maximum control bandwidth, we chose
to work with a Field Programmable Gate Array (FPGA) from Xilinx.

The logic structure of a Xilinx FPGA is designed to handle
arbitrary algorithm architectures. The FPGA mostly consists of a
grid with thousands of Configurable Logic Blocks (CLB) connected
by programmable interconnections. Each CLB contains a few small
look-up tables which can serve as a simple logic elements (AND,
OR, etc) when programmed. Also interspersed in this grid are
larger blocks of RAM that can be programmed as user defined
functions with a large domain and range. Since each logic element
needs to be triggered to operate, the distribution of a uniform
clock signal with constant frequency and phase is a considerable
design issue. Thus FPGA architectures commonly have digital clock
managers (DCM) or delay locked loops (DLL) that de-skew the clock
signal across the device.

The performance of FPGA architectures has been impressively increasing in
recent years. To give a current indication of their level of performance, we
quote some of the characteristics of one of the top of the line devices
available on the market today. The Xilinx Virtex II can contain up to 10
million system gates and have an internal clock frequency ($f_{C}$) up to 420
MHz. The input-output speed can be above 840 Mb/s which roughly matches the
maximum speed of the best analog to digital converters (100 MSPS for a 12 bit
sample Analog AD9432). This same FPGA has up to 192 SelectRAM blocks of 18
kbit each.  Because a strong demand from industry drives the development of
FPGA technology, these performance specifications will likely improve
significantly in the short term future.

Of course these devices must be coupled to a board, introducing other
practical issues. The system used in the cavity lock described below is a
GVA-290 board (G.V. \& Associates) with two Xilinx Virtex-E XCV1000E FPGA
chips. Signals enter and exit the board through four input and four output
SMA connectors. The signals are digitized by an ADC (Analog AD9432) at the
input and converted back to analog by a DAC (Analog AD9762) at the output.
Each ADC is located on a detachable daughter board, allowing for converter
upgrades and the addition of customized components and filters. Both the ADCs
and DACs have 12 bit resolution and are driven at the clock speed of 100 MHz.
A crystal oscillator provides the clock signal to the FPGA, which distributes
a synchronized signal internally with DLLs and also outputs the driving
signal for the ADC and DAC at a controlled phase.  Unlike standard models,
the board was ordered with DC coupled inputs, allowing us to have broadband
control to DC. Boards often come with anti-aliasing analog filters, but were
not included here due to the substantial group delay a high-order filter can
impose on the signal. The cost of this particular board including devices is
approximately \$10,000, but it should be stressed that functional systems
could be assembled at far less cost.

Xilinx also offers a special academic program through which
university researchers can obtain the necessary software
environment and a limited range of hardware products.

\begin{figure}
\includegraphics[width=3in]{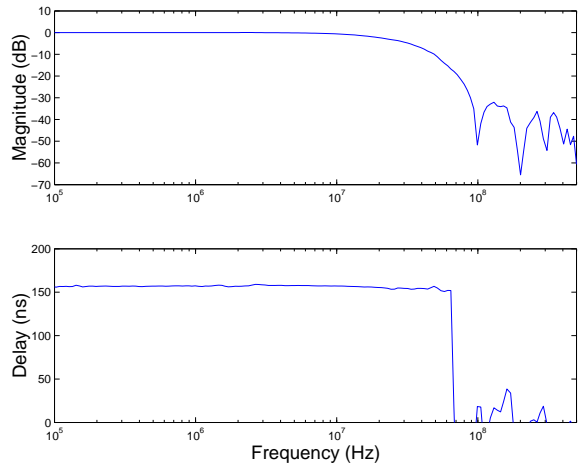}
\caption{The amplitude response and delay of the entire GV-290
board (ADC $\rightarrow$ FPGA $\rightarrow$ DAC). Notice that the
delay below the Nyquist frequency ($f_{C}/2=50$ MHz) is $\sim$ 160
ns. The phase response in the constant delay region is linear with
slope proportional to the delay. }
\label{identity}
\end{figure}

We can now discuss the latency and throughput of our controller in
more detail. The latency is defined as the amount of time for an
algorithm to process a single sample all the way through. The
throughput is defined as the number of samples (or bits) per
second being output from the device. For example, consider a
system of $N$ components in series, each with the same sampling
rate $f=\frac{1}{\tau}$. Also assume the system is `pipelined'
meaning that a new sample is loaded every $\tau$ seconds and
samples are registered (values held) in-between components.  In
this case, the latency is $N\tau$, while the throughput is $f$. If
this were a controller, the bandwidth of control would be limited
to the inverse of the latency $\frac{1}{N\tau}$, not the
throughput.

One of the principle advantages of FPGA technology is that the delay can be
quite small. Consider the case where the FPGA of the GVA-290 board is
programmed to pass a signal through without any manipulation. Figure
\ref{identity} shows the transfer function and delay of this configuration.
The ADC, FPGA, and DAC are all clocked at 100 MHz and each one takes a
certain number of cycles (10 ns/cycle) to perform its function.  The ADC
imposes a delay of 10 cycles, the buffers of the FPGA impose a delay of 4
cycles, and the DAC only delays the signal about 1 cycle.  Adding all this to
a small delay from other components, we find that below the Nyquist frequency
($f_{C}/2=50$ MHz) the signal passes through at unity gain with a constant
overall delay of $\sim$ 160 ns. Thus the maximum control bandwidth for this
device is $\sim$ 6 MHz, and bandwidths in the tens of MHz may be anticipated
with newer versions. If the FPGA algorithm is simple enough that the ADC
dominates the delay, it may be desirable to use Flash ADCs that have less
latency at the expense of a larger power consumption and smaller number of
output bits.

If the FPGA performs a complex calculation that requires multiple logical
steps in series, the delay is increased by an integer number of cycles and
the effective bandwidth suffers. A typical example is that of the FIR filter
mentioned below where, for $B_{U}$ input bits, the sampling rate becomes
$f_{C}/B_{U}$.  For any general algorithm, care should be taken to minimize
the number of serial elements before implementation.  If possible,
calculations should be performed in parallel and look-up tables should be
used to evaluate complicated functions.

\subsection{Software}

The design process for a particular algorithm has been largely automated with
implementation software environments like Foundation ISE (Xilinx). Once the
design is entered via one of the options described below, the program steps
through a series of compilation tasks before downloading onto the device. In
order, the design is analyzed for syntactic errors, synthesized into a
generic circuit, and implemented into an optimal bit stream appropriate to
the particular device and board. The bit stream is then downloaded onto the
device to achieve a stand-alone realization of the desired algorithm.
Simulation programs are available at intermediate stages for debugging
purposes. The latest version of Foundation ISE (4.1) compiles up to 100,000
gates/min. For reasonable designs, an entire design flow can be expected to
take about 10 minutes. This allows for a rapid prototyping cycle which is one
of the most desirable features of this technology.

Numerous algorithm entry options are available. Using a library of primitive
components, one can create a schematic of the desired circuit. Abstract
finite state machine diagrams can also be interpreted. The third option is a
text based design written in either Verilog or VHDL (VHSIC Hardware Design
Language).

As is common in technology standards, the choice of Verilog vs.\ VHDL has
become a religious one for everyday practitioners. It is worth pointing out
some of the accepted differences between the languages. Verilog is generally
regarded as being easier to learn. A strong majority of engineers
implementing commercial systems use Verilog. Historically, VHDL was meant as
a description language before being adopted as a means of synthesis. As a
result, VHDL is a much more strongly `typed' language. The range of
abstraction is also different between the two languages. Although there is a
considerable overlap, Verilog extends to a lower level of abstraction while
VHDL extends to a slightly higher level. For non-critical reasons, we chose
to design in VHDL, hence we will discuss the following designs in those
terms. However, the discussion is abstract enough that most concepts apply to
both languages.

To first order, VHDL is a text based description of a schematic design. The
mapping between input and output bus variables consists of a series of
abstractly defined components where output ports are connected to input ports
with defined signal variables. Each component has an associated `entity' and
`architecture', where an architecture is an instantiation of an entity. For
example, a component with entity `op-amp' (with only input and output ports
defined) could have its functionality determined by the particular
architecture `op27'. The internal workings of a particular architecture are
can be specified in another VHDL file with more components that are defined
elsewhere. In this way, the code lends itself nicely to nested level of
detail and organized project design. Also one can easily swap out components
by changing architectures, but not entities, within the code.

At some point in the hierarchy, primitive components must be called upon. The
Xilinx software offers an extensive library of such components (AND, OR,
etc.) for use with each particular device. In addition to these basic
primitives, one can also create more complicated, but commonly used,
components with the Xilinx `Core Generator'. These objects (adders,
multipliers, filters, DSP elements) can be customized with user specified
parameters.

Each component loads inputs and returns outputs triggered by an
input clock signal. Hence, when designing in VHDL one thinks in
terms of circuit diagrams where, on every clock cycle, events
happen concurrently across the device. On the other hand, in
traditional C-like computer languages events progress in a serial
manner. At times, serial logic is convenient and in fact VHDL
offers a restricted form of serial logic in a form known as a
`process'. These processes are bits of C-like code that execute
when triggered. Inside a process, variables can be manipulated
with functions defined in other VHDL files. However, a signal can
only be changed once within a process. For this and other reasons,
processes are best used as referees to generate secondary
triggering signals and logic. While processes can perform some
level of math, the heavy lifting is best left to the components
which have been streamlined for such purposes.

\begin{figure}
\includegraphics[width=2.6in]{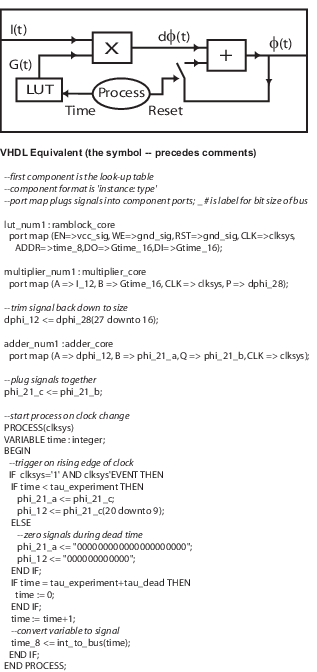}
\caption{FPGA schematic and corresponding code for the adaptive
phase measurement algorithm. In the schematic the process is not
represented as a block component because it is coded in a serial
manner.}
\label{adphi}
\end{figure}

\begin{figure}
\includegraphics[width=3in]{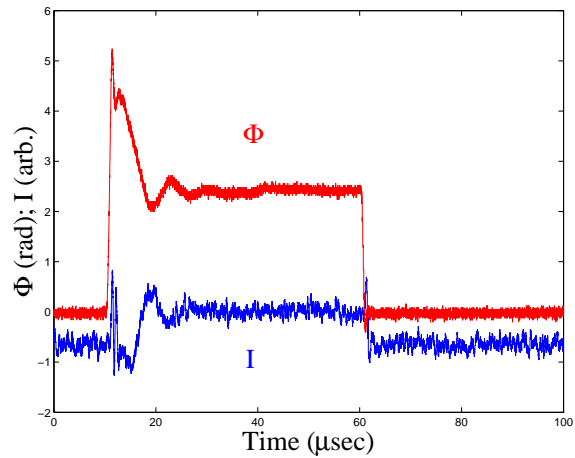}
\caption{The $\Phi(t)$ and $I(t)$ trajectories for the phase
measurement of a single pulse of light.  The current is locked to
zero and the ending point of the phase is a rough estimate of the
measured phase.  The true phase measurement is a functional of
both traces.  The small oscillations are due to the delay in the
loop. } \label{traj}
\end{figure}

An appropriate use of a process is to initialize parameters and control
timing. For example, Figure \ref{adphi} demonstrates how the simple adaptive
phase algorithm mentioned above is implemented. Both the VHDL and an
equivalent schematic are shown. The photocurrent, $I$, enters the device and
is multiplied by the time dependent gain factor, $G(t)=\frac{1}{\sqrt{t}}$,
which is created by sending the time signal, $t$, through a look-up table
(described below). The resulting signal, $d\Phi (t)=\frac{I(t)}{\sqrt{t}}$,
is then sent to one port of an adder, with the other input port being wired
to the output signal $\Phi (t)$. Because the output is connected to the input
with a delay, the adder serves as an integrator and executes the relation
$\Phi (t)=\Phi (t-1)+d\Phi (t)$ at every time step. The `process' plays an
important role in this algorithm by initializing the integral value and
creating the time signal. At the beginning of the pulse (integration), the
process initializes $t$ and $\Phi $ to zero. Every subsequent clock signal,
the process increments $t$ by one and lets the adder integrate up the signal.
At the end of the pulse, the process waits for the next pulse then repeats
the sequence.  Figure \ref{traj} shows the algorithm in action. Through the
integrator structure, $\Phi$ is adjusted until $I$ is locked to zero.  The
overshoot is a result of the FPGA delay.

A single measurement using this algorithm is shown in Figure \ref{traj}.
Here the `pulse' is a 50 $\mu$sec long time slice of a weak cw coherent beam.
The feedback algorithm is sampling at 100 MHz with a delay less than 1 $\mu$
sec. Because of the delay and other bandwidth limiting components in the
loop, our effective feedback bandwidth is limited to $\sim$ 1 MHz.

As will be demonstrated below, Matlab plays a complementary role in the
design process. It can be used to create the necessary coefficients and
memory blocks used as parameters in the VHDL components. In particular, the
Control and DSP toolboxes provide relevant functionality. Also, Simulink is a
good tool for simulating the associated experiments, where delays and other
realistic factors can complicate the dynamics. There exist software packages
that attempt to directly translate from a Simulink design of an algorithm
into equivalent VHDL, but these packages remain in early stages of
development.

Due to their extensive utility, RAM look-up tables and filter components are
worth discussing in greater detail.

\subsubsection{Look-Up Tables}

Most FPGA chips come equipped with large blocks of internal RAM that can be
used as generalized functions or look-up tables (LUT). Given an amount of
memory on a particular block, the user can decide on a certain number of
input and output bits. During operation the RAM block returns the value held
at the address specified by the input, effectively implementing the desired
function. For example, on the XCV1000E, 160 blocks of 4096 = $2^{12}$ bits
are available for internal use. (As noted above, the Virtex II devices have
much larger 18 kbit blocks.) To make one block behave as the function $f$
with $B_{i}$ input bits, the designer would choose the output to be
$B_{o}=2^{12-B_{i}}$ bits. Possible partitions are
$(B_{i},B_{o})\in[(1,2048),(2,1024),(3,512)...,(8,16),...(12,1)]$. Once a
partition is chosen, the designer would use Matlab to define a block of data
consisting of $2^{B_{i}}$ values each of size $B_{o}$ bits, and use this
block of data as a parameter in the VHDL LUT component. If the discretization
is a problem, more RAM blocks can be used to represent the function. If
desired, the memory of a RAM block can also be dynamically written during
operation.  With this ability, an algorithm could easily adapt itself
according to the signals it receives. Both the read and write operations
(from/to one RAM address) only take a single clock cycle.

As mentioned above, these LUT functions play an extremely important role in
speeding the functionality of non-linear algorithms. The application may be
as simple as non-linear gain-scheduling of a controller or as complicated as
full quantum-mechanical state estimation with the LUT performing functions
based on assumed system parameters. In general, it is a matter of judgment
how to partition complex algorithms, but any optimal partition will likely
involve the use of these LUTs to perform the difficult parts of the
calculation with minimal time delay.

\subsubsection{Filters}

PLD's have a clear edge over analog circuitry in non-linear processing, but
they also have a potential advantage in implementing precise, generic linear
filters and transfer functions.

A standard core element offered by Xilinx is the FIR (Finite Impulse
Response) filter. The FIR is defined in discrete time as
\begin{equation}
y(n)=\sum_{i=0}^{N}a(i)u(n-i)
\end{equation}
where $y(n)$ and $u(n)$ are the output and input at the discrete time $n$
respectively. With standard Matlab functions (\texttt{firls}, \texttt{remez})
one can specify an arbitrary amplitude response and get out the corresponding
$a(i)$ vector. The sampling frequency for a FIR element is $f_{F}=
\frac{f_{C}}{B_{U}}=\frac{1}{\tau_{F}}$ where $B_{U}$ is the number of bits
chosen to represent $u(n)$. Of course, the filter is useless at shaping the
response above this frequency. The group delay of the signal through the
filter is approximately $\tau_{F}\frac{N}{2}$.

The range of attenuation is also a concern in the design of any filter. For
an FPGA with $B_{F}$ bits entering and leaving, the dynamic range is $20\log
(2^{B_{F}})$dB. For our board with 12 bit ADC/DAC inputs and outputs, this
corresponds to 70 dB. The designer should also have a sense of the size of
the input and output signals. If the input signal is too high, the FPGA will
rail; if the input is too low, it will fail to rise above the smallest bit
size. To avoid these types of problems, broadband gain elements can be used
at the input and output of the FPGA board.

A drawback of the FIR design is that it cannot be used to control the phase
response of its transfer function. On the other hand, a generic continous
time linear transfer function
\begin{equation}
G_{C}(s)=\frac{c(N)s^{N}+c(N-1)s^{N-1}+...+c(1)}{%
d(N)s^{N}+d(N-1)s^{N-1}+...+d(1)}
\end{equation}
where $Y_{C}=G_{C}U_{C}$, has phase control built in through the denominator.
To approximate this function on a PLD, an Infinite Impulse Response (IIR)
filter needs to be used.

\begin{figure}
\includegraphics[width=3in]{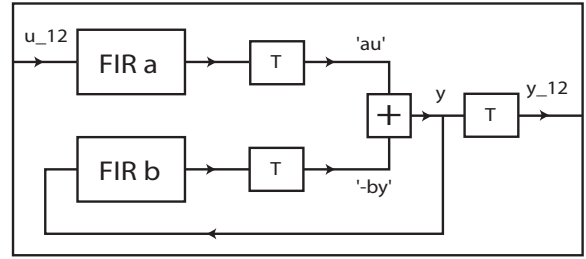} \caption{Implementation of IIR filter.
`T' components trim a certain number of least significant bits
from the data bus. }\label{iir}
\end{figure}

One possible IIR design process illustrates this need.  To generate a digital
IIR design, first create $G_{C}(s)$ using standard control techniques
(Nyquist, LQR, etc.).  Next, convert from a continuous to a discrete transfer
function
\begin{equation}
G_{C}\Rightarrow G_{D}(z)=\frac{a(0)+a(1)z^{-1}+...+a(N)z^{-N}}{%
b(0)+b(1)z^{-1}+...+b(N)z^{-N}}
\end{equation}
with the Matlab function \texttt{c2d}.  We have used the definition
$Y_{D}=G_{D}U_{D}$ in the discrete time representation. Apply a z-transform
($z^{-1}\Rightarrow $ unit delay) to create the discrete time difference
equation
\begin{equation}
y(n)=\sum_{i=0}^{N}a(i)u(n-i)-\sum_{i=1}^{N}b(i)y(n-i)
\end{equation}
with the definition $b(0)=1$. Finally, implement the difference
equation in hardware as in Figure \ref{iir} with 2 FIR blocks and
1 adder.

With $b(n>0)=0$ the filter is just a FIR filter, however with
$b(n>0)\neq 0$ the output is fed back to itself. Hence an impulse
response will have an infinite effect on the output. Of course,
with internal feedback loops, the system is potentially unstable
to noise and rounding errors. For this reason, among others, the
Xilinx `Core Generator' does not create flexible IIR modules.

However, with careful consideration of the number of bits required at each
stage, a stable IIR filter can be created as in Figure \ref{iir}. The
sampling frequency for this simple architecture is $\frac{f_{C}}{2 B_{Y}}$
where $B_{Y}$ is the number of bits used to keep track of $y(n)$ internally.
The factor of 2 results from the delay of both the adder and the FIR
element.\ Because of the feedback, the IIR filter can achieve a given
amplitude response with lower number of coefficients than the FIR filter.
This means the filter delays the signal less.  Even though the IIR has fewer
coefficients than an analagous FIR filter, the coefficients of the IIR filter
have to be specified to a greater degree of precision to achieve the same
amplitude response.

\section{Specific Example: Cavity Lock}

We now discuss the use of an FPGA to perform a classical task necessary for
low-noise experiments. High precision optical measurements demand laser
intensity noise be minimized as much as possible. In the adaptive phase
experiment mentioned above, the input laser is a Lightwave Nd:YAG model 126
(1064 nm) with an inherent broad relaxation oscillation noise peak at $\sim$
100 kHz. \ To perform broadband detection and control near 1 MHz, this
intensity noise must be removed from the beam with a Fabry-Perot cavity.

\begin{figure}[t]
\includegraphics[width=3in]{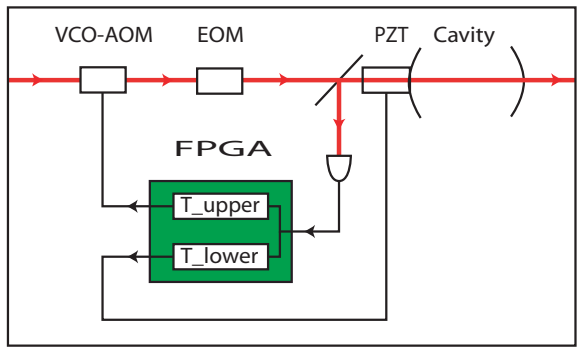}
\caption{Feedback architecture for a Fabry-Perot Cavity. The EOM
puts sidebands on the beam necessary to generate the locking
signal. The FPGA algorithm T\_upper maps the error signal to the
fast VCO-AOM frequency shifting combination. The FPGA algorithm
T\_lower maps the signal to the slow PZT. }
\label{cavlock}
\end{figure}

A block diagram of the system is shown in Figure \ref{cavlock}. The output
intensity of the cavity is stabilized with the standard Pound-Drever-Hall
method so that the error signal is created from a reflected carrier beam with
sidebands. At low frequencies (below 100 Hz) the feedback loop is dominated
by a piezoelectric element (PZT) which controls the length of the cavity. At
higher frequencies and through the closing point of the servo, the feedback
is from an AOM (Acousto-Optic Modulator) driven by a VCO (Voltage Controlled
Oscillator) which adjusts the frequency of the input beam.

Given the control architecture of Figure \ref{cavlock}, the design process
can be made very systematic with the flexibility of the FPGA. Because the
critical behavior of the servo will be dominated by the VCO-AOM loop, we
concentrate on the design of $T_{U}$ (T\_upper).  First, the transfer
functions of the elements in the loop are measured.  Here we find that the
VCO-AOM combination behaves like a low-pass filter ($T_{V}$) with a corner at
100 kHz. The cavity itself can be modelled as a low-pass filter ($T_{C}$)
with a corner at about 10 kHz (the cavity linewidth).  The goal is to design
$T_{U}$ such that the closed loop transfer function
$T_{CL}=\frac{T_{C}T_{V}}{1+T_{C}T_{V}T_{U}}$ is stable.

At this point, we can use the Matlab Control Toolbox to design an optimal
$T_{U}$.  One option is to provide the function $\texttt{lqr}$ with the state
space representations of $T_{V}$ and $T_{C}$ and an appropriate cost function
to create the optimal $T_{U}$.  The result simply tells us to make the
combination $T_{C}T_{V}T_{U}$ behave like an integrator
($T_{I}=\frac{1}{s}=\frac{1}{j\omega}$) such that the controller satisfies
the Nyquist criterion with 90 degrees of phase margin.

There are practical problems with this approach. In particular, the gain of
$T_{U}$ must be infinite for very low and very high frequencies.  To remedy
this, we flatten the response of $T_{U}$ below 100 Hz (where the PZT arm
takes over) and roll off the response at 300 kHz, beyond the closing point of
the servo. So instead of making $T_{U}=\frac{T_{I}}{T_{C}T_{V}}$ we use
$T_{U}=\frac{T_{LP1}T_{LP2}^{2}}{T_{C}T_{V}}$ where $T_{LP1}$ is a low-pass
filter with the corner at 100 Hz and $T_{LP2}$ is a low-pass filter with the
corner at 300 kHz.

To get high gain at frequencies below 100 Hz, we make $T_{L}$ (T\_lower)
behave as a low-pass filter with a corner at only a few Hz.  A better choice
would be to implement $T_{L}$ as a high-gain analog integrator, but we use
the FPGA to implement $T_{L}$ here for demonstration purposes.

Next, we generate proper IIR coefficients for both paths by the method
described previously, treating $T_{L}$ and $T_{U}$ as the continuous transfer
function $G_{C}$.  With a clock frequency of 100 MHz and an internal sample
size of $B_{Y}=32$ bits, the IIR structure had an effective bandwidth of 1.5
MHz ($\frac{f_{C}}{2 B_{Y}}$), which is adequate to generate the critical
features of the transfer function.

\begin{figure}[top]
\includegraphics[width=3in]{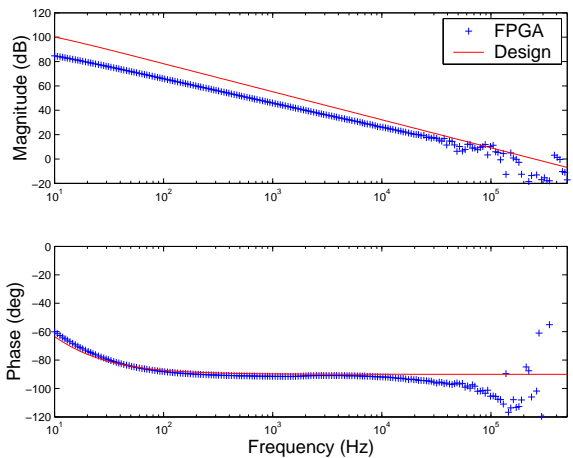}
\caption{Bode plot of T\_lower (transfer function leading to PZT).
The design is a low-pass filter which dominates control below
$\sim$ 100 Hz. } \label{tf_fpga_lo}
\end{figure}

\begin{figure}[top]
\includegraphics[width=3in]{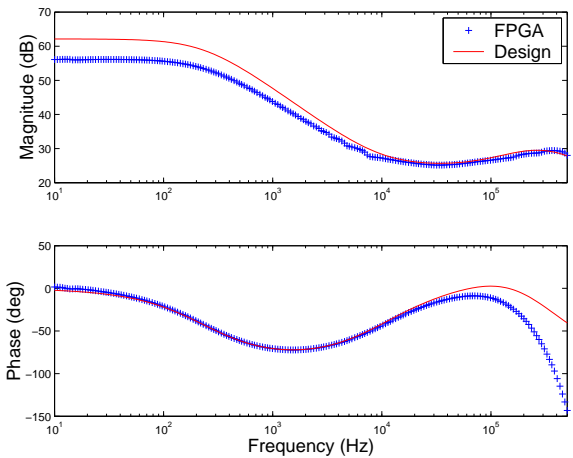} \caption{
Bode plot of T\_upper (transfer function leading to VCO-AOM).  The
peak in phase is designed to stabilize the plant through the unity
gain point. }\label{tf_fpga_up}
\end{figure}

Figures \ref{tf_fpga_lo} and \ref{tf_fpga_up} show the desired and
actual transfer functions of both arms.  Each arm fails to match
the desired phase and amplitude response in a similar way. First,
because of the finite size of the sampling time, the actual phase
response differs from the desired response as the frequency
approaches the effective sampling frequency. In fact, this
mismatch happens lower than the sampling frequency because of the
delay of the IIR filter. Second, at low frequencies, the FPGA
gives less gain than the desired result. This is due to the fact
that we are dealing with finite precision coefficients. The price
paid for having a large sampling frequency with small delay is
that we have less control over the size of the low frequency gain.
Finally, note that the PZT arm integrator achieves the full 70dB
of expected range (input/output size is 12 bits).

The closed loop transfer function behavior for both arms matches our
expectations for noise rejection at low frequencies. A mismatch at higher
frequencies is due to inadequate modelling of the PZT and other components.
(The PZT behaves more like a collection of oscillators with different
resonances than a low pass filter.) Qualitatively, the FPGA lock was much
more robust to high frequency noise than an analog version of the servo. This
was likely due to the precise match to the plant dynamics near the unity gain
point of the servo, achieved by the use of large FIR coefficients. However,
the FPGA lock was unable to retain the lock over time scales more than a few
hours due to the saturated gain at very low frequencies.  This problem could
easily be remedied by using an analog integrator with more DC gain to replace
the FPGA PZT transfer function.  The main advantage of the FPGA is its fast
accurate response and, besides the demonstration presented here, there is no
practical reason to use the FPGA for high-gain, low-frequency applications.

Finally, another feature of FPGA control is the possibility of adding logical
automation to this system. Specifically, if the controller loses the lock,
then the FPGA could be programmed to sense this condition, sweep for a
signal, hone in, and re-acquire the lock. The abstract logical nature of VHDL
code makes this task simple relative to the procedure needed to create an
acquisition system using standard electronics.

\bigskip

\section{Summary}

To demonstrate the use of programmable logic technology in an otherwise
familiar setting, we have concentrated on a linear control application. We
have used this example to convey the issues associated with a digital
controller, including design, latency, and discretization.  However, we have
only hinted at the more interesting advanced applications in experimental
quantum optics which are sure to develop more quickly because of this
technology. FPGAs and similar devices are particularly suited to any physical
system where non-linear mappings are desired between output and input
variables within the natural dynamical time-scale.  With these devices and
sufficiently protected quantum systems in hand, the field of coherent quantum
control may soon have enough speed to match the intelligence of its proposed
controllers.

\bigskip
\textbf{ACKNOWLEDGEMENTS}
\bigskip

J.S. acknowledges the support of a Hertz Foundation Fellowship,
and H.M. acknowledges the support of an A. P. Sloan Research
Fellowship. This work was supported by the NSF under grant PHY-9987541, and by the
ONR under Young Investigator Award N00014-00-1-0479.

\end{document}